\documentclass[12pt]{iopart}

\usepackage[latin1]{inputenc}%
\usepackage[T1]{fontenc}     %
\usepackage{iopams}          %
\usepackage{graphicx}        %

\eqnobysec

\newcommand{\tab}{\raisebox{-1.7ex}[0pt][0pt]} 

\begin{document}

\title[Generalized Laguerre Polynomials with PDEM and Wigner's Functions]{Generalized Laguerre Polynomials with Position-Dependent Effective Mass Visualized via Wigner's Distribution Functions}

\author{O Cherroud\footnote{Permanent address: Faculté de technologie, université Yahia FAR\'ES de Médéa, Médéa 26\,000, Algeria}, S-A Yahiaoui and M Bentaiba}

\address{LPTHIRM, Département de physique, Faculté des sciences, Université Saâd DAHLAB-Blida~1, B.P.~270 Route de Soumâa, 09\,000 Blida, Algeria}
\eads{\mailto{Cherroud.Othmane@univ-medea.dz},\\ \qquad\quad\mailto{s$\_$yahiaoui@univ-blida.dz},\\ \qquad\quad\mailto{bentaiba@univ-blida.dz}}

\begin{abstract}
We construct, analytically and numerically, the Wigner distribution functions for the exact solutions of position-dependent effective mass Schr\"odinger equation for two cases belonging to the generalized Laguerre polynomials. Using a suitable quantum canonical transformation, expectation values of position and momentum operators can be obtained analytically in order to verify the universality of the Heisenberg's uncertainty principle.
\end{abstract}


\section{Introduction}%

\noindent Schr\"odinger equation (SE) endowed with position-dependent effective mass (PDEM) has received a growing interest on behalf of physicists during the last decade \cite{1,2,3,4,5,6,7,8,9,10,11,12}. Its solutions has been found to be very useful in describing, physically, the properties of the quantum dynamics of electrons in condensed matter physics as well as related fields of physics \cite{13,14,15,16}. In mathematical physics, they have found to be interesting in point of view of coherent states \cite{17,18,19,20} and $\mathcal{PT}$-symmetry \cite{21,22,23}. A lot of methods and approaches, including the factorization method \cite{24}, supersymmetry of quantum mechanics and the related shape-invariant potentials \cite{25}, Lie algebra \cite{26,27,28}, path integral \cite{29} and operators techniques \cite{30}, have been applied to the system with PDEM to obtain algebraically the exact solutions. The coordinate transformation (CT) is one of these methods and there are two kinds (see, for example \cite{31} and references therein). The first one connects two different solvable potentials, while the second one allow us to convert SE into the second order differential equation which has solutions of the special functions such as (confluent-)hypergeometric functions and orthogonal polynomials.\\
\indent However in quantum mechanics, as is well known, we are faced with the measurability of position ($\hat x$) and momentum ($\hat p$) operators of the quantum state imposed by the Heisenberg's uncertainty principle, which no longer holds in classical mechanics. Then in order to restor the connection between quantum and classical mechanics, it would be desirable to work over the phase space and to have some functions that display the ordinary $c$-number coordinates $x$ and $p$ equally within the desired model. The Wigner's quasi-probability distribution in conjunction with the Weyl transformation are functions in which we are looking for \cite{32,33,34,35,36,37}. Besides the conventional Hilbert space (Heisenberg, Schr\"odinger and Dirac) formulation of quantum mechanics and path integral (Feynman) approach, the Wigner distribution functions (WDF) furnish an alternative treatment for describing quantum mechanics, independently of the first two formalisms. See also \cite{38,39,40,41,42,43,44,45,46,47} where some applications of quantum systems with Wigner functions and Weyl transformation were studied. \\
\indent By definition, given the eigenfunctions $\psi_n(x)$, WDF in the phase space is given by
\begin{eqnarray}\label{1.1}
  \mathcal W(\psi_n|x,p)&=&\frac{1}{2\pi}\int_{-\infty}^{+\infty}\rme^{-\rmi py}\,\psi^\ast_n\left(x-\frac{\hbar y}{2}\right)\psi_n\left(x+\frac{\hbar y}{2}\right)\,\rmd y,
\end{eqnarray}
and satisfying the following properties: (i) reality: $\mathcal W(\psi_n|x,p)=\mathcal W^\ast(\psi_n|x,p)$, (ii) position probability density: $\mathfrak P(x)=\int\mathcal W(\psi_n|x,p)\,\rmd p$, (iii) momentum probability density: $\mathfrak P(p)=\int\mathcal W(\psi_n|x,p)\,\rmd x$, (iv) normalization: $\int\!\!\int\mathcal W(\psi_n|x,p)\,\rmd x\,\rmd p=1$, and (v) bounded: $|\mathcal W(\psi_n|x,p)|\leq1/(\pi\hbar)$.\\
\indent Using \eref{1.1} the expectation value of an operator $\hat{\mathcal O}$ is given through
\begin{eqnarray}\label{1.2}
  \langle{\mathcal O}\rangle&=&\int_{-\infty}^{+\infty}\rmd x\int_{-\infty}^{+\infty}\rmd p\,\,\mathcal W(\psi_n|x,p)\,\mathfrak{W}[\hat{\mathcal O}],
\end{eqnarray}
where $\mathfrak{W}[\hat{\mathcal O}]$ is called the Weyl transformation of the operator $\hat{\mathcal O}$. It is the central object in the phase space (or deformed-) quantization and is defined by
\begin{eqnarray}\label{1.3}
  \mathfrak{W}[\hat{\mathcal O}]&=&\hbar\int_{-\infty}^{+\infty}\rme^{-\rmi py}\,\left\langle x+\frac{\hbar y}{2}\left|\hat{\mathcal O}\right|x-\frac{\hbar y}{2}\right\rangle\,\rmd y.
\end{eqnarray}
\indent Aside from the WDF calculated for a system with a constant mass which have been well understood in many works, comparatively little work, to our knowledge, has been done on WDF for PDEM \cite{48,49}, whence the purpose of this paper is to fill this gap. Using the coordinate transformations (CT), the eigenfunction of many quantum systems can be expressed in terms of orthogonal polynomials. In this paper we mainly calculate, analytically and numerically, WDF in the case of the generalized Laguerre polynomials (GLP) \cite{31}, using to that end an exponentially decaying mass function $m(x)=m_0\,\rme^{-\alpha|x|}$ \cite{31,50}, in hopes to see how the mass function can affect WDF. The use of quantum canonical transformations $(\hat x,\hat p)\rightarrow\left(\mu(\hat x),\pi(\hat x,\hat p)\right)$ \cite{51} will allow us to calculate the expectation values of the new coordinates $(\mu,\pi)$ in order to reproduce the Heisenberg uncertainty principle. Our analysis reveals that: (i) WDF have a typical triangular shapes different from those studied in \cite{48,49}, (ii) WDF are compressed in the $x$-direction and become stretched in the direction of momentum $p$ when the quantum number $l$ increases, and (iii) in the limit $l\rightarrow\infty$ we can observe that a common pattern emerges, for the uncertainty principle, as a linear-behavior
\[
  (\Delta\mu)_n\cdot(\Delta\pi)_n\simeq n+\frac{1}{2},
\]
applied to all cases and levels $n=0,1,2,\cdots$\\
\indent This paper is organized as follows: in the next section we start by describing a brief introduction of CT and present their exact solutions for PDEM SE in the case of GLP. In section 3 we find analytically WDF for two cases of Laguerre PDEM and we plot their distributions containing some features. In both cases we consider an exponentially decaying mass function. Quantum canonical transformation is introduced in section 4 to find a correspondence between classical and quantum variables involving in our problem. This is followed by evaluating a spread in position and momentum in order to verify the Heisenberg uncertainty principle. Finally the last section is devoted to our comments and conclusion. Two appendices are added in order to complete some details and proofs omitted in the main text.

\section{Coordinate transformation and Laguerre exact solutions of PDEM SE}%

\noindent Taking the natural units ($\hbar=m_0=1$) and using the ordering prescription adopted by BenDaniel and Duke \cite{52}, the one-dimensional PDEM SE can be expressed as
\begin{eqnarray}\label{2.1}
  \left(-\frac{1}{2}\frac{\rmd^2}{\rmd x^2}+\frac{m'(x)}{2m(x)}\frac{\rmd}{\rmd x}+m(x)V(x)\right)\psi(x)=m(x)E_n\psi(x).
\end{eqnarray}
\indent Then by applying the following CT, $\psi(x)=f(x)F\big(g(x)\big)$, to the eigenfunctions, it is not difficult to verify that \eref{2.1} satisfies the second order differential equation \cite{31}
\begin{eqnarray}\label{2.2}
  \frac{\rmd^2F(g)}{\rmd g^2}+Q(g)\frac{\rmd F(g)}{\rmd g}+R(g)F(g)=0,
\end{eqnarray}
where $F(g)$ is some special function on $g(x)$. The functions $Q(g)$ and $R(g)$ are given by
\begin{eqnarray}
  Q(g) &=& \frac{g''(x)}{g'^2(x)}+\frac{2f'(x)}{f(x)g'(x)}-\frac{m'(x)}{m(x)g'(x)},\label{2.3} \\
  R(g) &=& \frac{f''(x)}{f(x)g'^2(x)}-\frac{m'(x)f'(x)}{m(x)f(x)g'^2(x)}+\frac{2m(x)}{g'^2(x)}\,\big(E_n-V(x)\big).\label{2.4}
\end{eqnarray}
\indent Integrating \eref{2.3}, we arrive to express $f(x)$ as
\begin{eqnarray}\label{2.5}
  f(x) &=& \sqrt{\frac{m(x)}{g'(x)}}\,\exp\bigg\{\frac{1}{2}\int^{g(x)}Q(g)\,\rmd g\bigg\},
\end{eqnarray}
and by inserting \eref{2.5} into \eref{2.4}, one can see that we obtain a system where the associated effective potential depends on the mass function
\begin{eqnarray}\label{2.6}
\fl  E_n-V_{\rm eff}(x) &=& \frac{g'^2(x)}{2m(x)}\,\left(R(g)-\frac{1}{2}\frac{\rmd Q(g)}{\rmd g}-\frac{1}{4}Q^2(g)\right)+\frac{1}{4m(x)}\,\big(S(g')-S(m)\big),
\end{eqnarray}
where $S(z)=z''/z-3/2\,(z'/z)^2$ is Schwartz's derivative of the function $z(x)$ and the prime denotes the derivative with respect to $x$. It follows that the PDEM SE can be solved if the forms of $Q$ and $R$ are given for a mass function $m(x)$. In order to obtain the effective potential in the above equation, we impose that there must be a constant on the right-hand side of \eref{2.6} representing the bound-state energy spectrum $E_n$ on the left-hand side.\\
\indent From \eref{2.5} the solution of the eigenfunctions $\psi_n(x)$ are given by
\begin{eqnarray}\label{2.7}
  \psi_n(x)&\sim &\sqrt{\frac{m(x)}{g'(x)}}\,\exp\bigg\{\frac{1}{2}\int^{g(x)}Q(g)\,\rmd g\bigg\}\,F_n\big(g(x)\big),
\end{eqnarray}
up to a normalization constant. It is worth to note that all expressions reduce to the well known ones if the mass is taken to be constant, i.e. $m(x)=1$.\\
\indent In the remainder of the paper, we choose to work under the special function $F_n(x)$ to be the generalized Laguerre polynomials $L_n^{(a)}(x)$. Here the functions $Q(g)$ and $R(g)$ are defined through
\begin{eqnarray}\label{2.8}
  Q(g)=\frac{a+1}{g(x)}-1,\qquad R(g)=\frac{n}{g(x)},
\end{eqnarray}
where $n$ is non negative integer and $a\neq-m,\,\,(m\in\mathbb N^\ast)$. Substituting \eref{2.8} into \eref{2.6} we arrive at the equation
\begin{eqnarray}\label{2.9}
\fl  E_n-V_{\rm eff}(x)=\frac{g'^2(x)}{4m(x)g(x)}(2n+a+1)+\frac{g'^2(x)}{2m(x)g^2(x)}\left(\frac{a+1}{2}-\frac{(a+1)^2}{4}\right)-\frac{g'^2(x)}{8m(x)}\nonumber\\
          +\frac{1}{4m(x)}\left(S(g')-S(m)\right).
\end{eqnarray}
\indent More generally, equation \eref{2.9} can be solved by choosing an appropriate $g(x)$ in order to make the right-hand side have a constant dependent on $n$. In doing so, we distinguish three different cases studied in \cite{31} and we refer them here by LI, LII and LIII, where each case has its appropriate $g(x)$ function. At this point, we are able to study some particular cases and we will focus our attention on the cases LI and LIII.

\paragraph{Case LI.}%
\noindent According to \cite{31}, when choosing that $g(x)$ satisfies the differential equation $g'^2(x)=4\,\omega\, m(x)\,g(x)\,\,(\omega>0)$, then the effective potential $V_{\rm eff}^{(I)}(x)$, the energy eigenvalues $E^{(I)}_n$ and the eigenfunction $\psi^{(I)}_n(x)$ to the case LI are given by
\begin{eqnarray}
\fl  V_{\rm eff}^{(I)}(x) &=& -\left(l+\frac{3}{2}\right)\omega+\frac{\omega^2}{2}\mu^2(x)+\frac{l(l+1)}{2\,\mu^2(x)}
                 +\frac{1}{8m(x)}\left[\frac{m''(x)}{m(x)}-\frac{7}{4}\left(\frac{m'(x)}{m(x)}\right)^2\right],\label{2.10} \\
\fl  E^{(I)}_n &=& 2n\,\omega,\label{2.11} \\
\fl  \psi^{(I)}_n(x) &=& \mathcal N_n^{(I)}\,m^{1/4}(x)\,\mu^{l+1}(x)\exp\left\{-\frac{\omega}{2}\,\mu^2(x)\right\}\,L_n^{(l+1/2)}\left(\omega\mu^2(x)\right),\label{2.12}
\end{eqnarray}
where $l=a-\frac{1}{2}\,\,\left(l\neq-\frac{3}{2},-\frac{5}{2},-\frac{7}{2},\cdots\right)$ and for convenience we have introduced the auxiliary mass function
\begin{eqnarray}\label{2.13}
\mu(x):=\int^x\sqrt{m(\eta)}\,\rmd\eta.
\end{eqnarray}

\paragraph{Case LIII.}%
\noindent If $g(x)$ satisfies the differential equation $g'^2(x)=4\,\omega^2\,m(x)\,\,(\omega>0)$, then $V_{\rm eff}^{(III)}(x)$, $E^{(III)}_n$ and $\psi^{(III)}_n(x)$ for the case LIII are as follows
\begin{eqnarray}
\fl  V_{\rm eff}^{(III)}(x) = \frac{b^2}{2(l+1)^2}-\frac{b}{\mu(x)}+\frac{l(l+1)}{2\,\mu^2(x)}
                 +\frac{1}{8m(x)}\left[\frac{m''(x)}{m(x)}-\frac{7}{4}\left(\frac{m'(x)}{m(x)}\right)^2\right],\label{2.14} \\
\fl  E^{(III)}_n = \frac{b^2}{2(l+1)^2}-\frac{b^2}{2(n+l+1)^2},\label{2.15} \\
\fl  \psi^{(III)}_n(x) = \mathcal N_n^{(III)}\,m^{1/4}(x)\,\mu^{l+1}(x)\exp\left\{-\frac{b}{n+l+1}\mu(x)\right\}\,L_n^{(2l+1)}\left(\frac{2b}{n+l+1}\mu(x)\right),\label{2.16}
\end{eqnarray}
where $\omega=b/(n+l+1)$ and $a=2l+1\,\,\left(l\neq-1,-\frac{3}{2},-2,-\frac{5}{2},\cdots\right)$.

\section{Wigner distribution functions for the generalized Laguerre PDEM}%

\noindent Now that the concepts of CT have been established for the cases LI and LIII, we will calculate the eigenfunctions given in \eref{2.12} and \eref{2.16} using an exponentially decaying mass function $m(x)=\rme^{-\alpha|x|}$, where $\alpha\sim 1/L$ being the inverse quantum-well width \cite{31,50}. According to \eref{2.13} the auxiliary mass function $\mu(x)$, for $\alpha\neq0$, is given by
\begin{equation}\label{3.1}
\mu(x)=\cases{
              -\frac{2}{\alpha}\,\rme^{-\alpha x/2},&(for $x > 0)$\\
              +\frac{2}{\alpha}\,\rme^{+\alpha x/2},&(for $x < 0)$\\
             }
\end{equation}
\indent Inserting \eref{3.1} into \eref{2.12} and \eref{2.16}, we get
\begin{eqnarray}
  \psi^{(I)}_n(x) &=& \mathcal N_{n,l}^{(I)}\,\mu^{l+3/2}(x)\,\exp\left\{-\frac{1}{2}\,\mu^2(x)\right\}\,L_n^{(l+1/2)}\left(\mu^2(x)\right),\label{3.2} \\
  \psi^{(III)}_n(x) &=& \mathcal N_{n,l}^{(III)}\,(2\mu(x))^{l+3/2}\,\exp\left\{-\mu(x)\right\}\,L_n^{(2l+1)}\left(2\mu(x)\right), \label{3.3}
\end{eqnarray}
where for convenience we also set $\omega=1$, (i.e., $b=n+l+1$ in the case LIII). Here the normalized constants are given, respectively, by
\[
\mathcal N_{n,l}^{(I)}=\sqrt{\frac{\alpha\,n!}{\Gamma\left(n+l+\frac{3}{2}\right)}},\quad\textrm{and}\quad \mathcal N_{n,l}^{(III)}=\sqrt{\frac{\alpha\,n!}{4(n+l+1)\Gamma(n+2l+2)}}.
\]
\begin{figure}
 \centering
 \includegraphics[width=15.5cm,height=8cm]{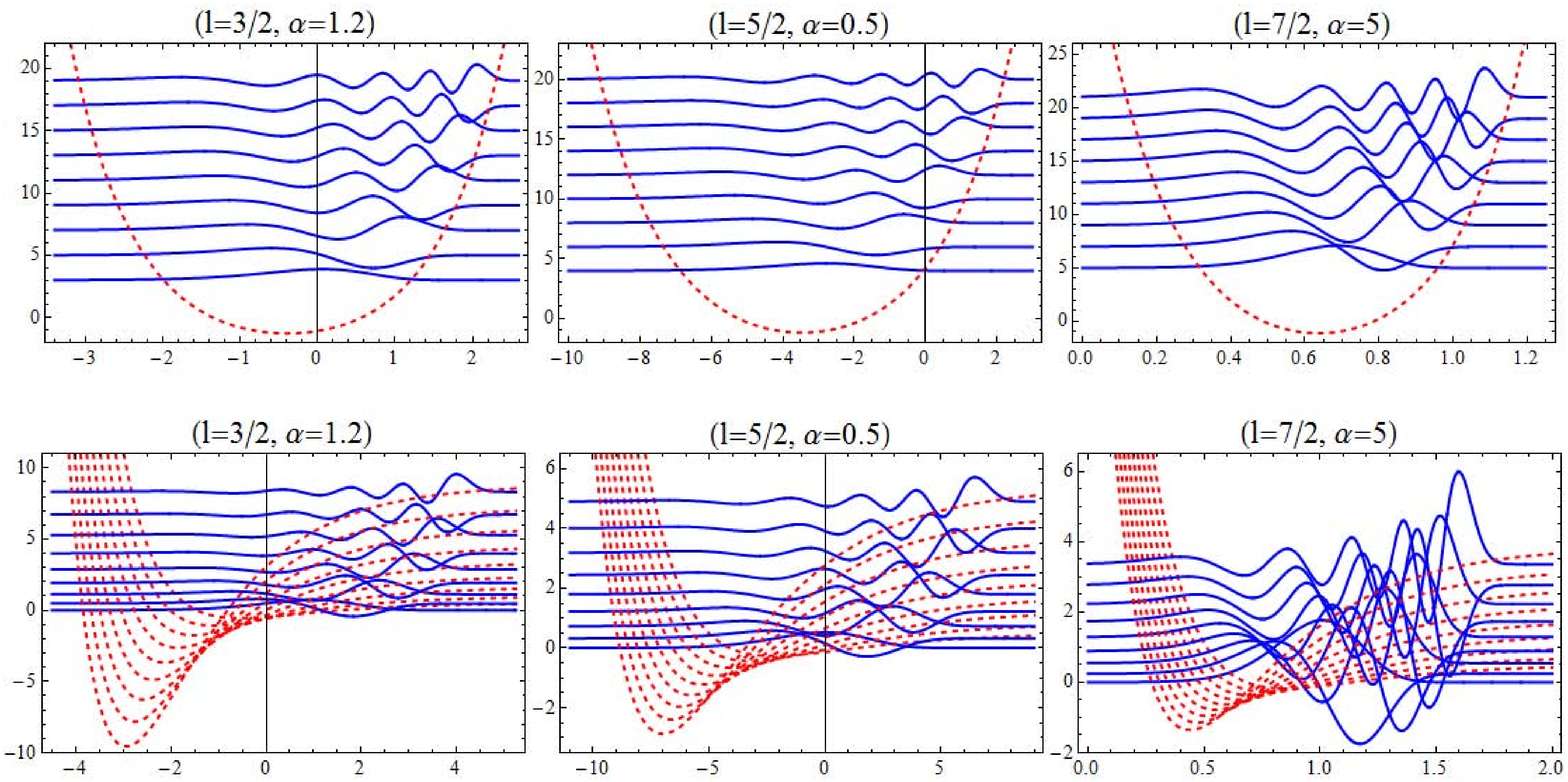}\\
 \caption{The eigenfunctions (blue lines) and their associated potentials (red lines) for even and odd $n$ up to 8. From top to bottom, the figures correspond to the cases LI and LIII, respectively. They are plotted for the mass function \eref{3.1} with different values of the parameters $\alpha$ and $l$. The horizontal eigenfunctions lines starting from the left also indicate the exact values of the corresponding energy levels $E_n^{(I,III)}$.} \label{figure1}
\end{figure}

\indent The characteristic curves for the deduced eigenfunctions \eref{3.2} and \eref{3.3} and their associated effective potentials are depicted in \Fref{figure1}. They are plotted for the effective mass function \eref{3.1} for even and odd quantum number $n$ up to 8, with different values of the parameter $\alpha=0.5,1.2,5$, and semi-integer values $l=\frac{3}{2},\frac{5}{2},\frac{7}{2}$. It is worth to note that the integer values for $l$ correspond to the three-dimensional harmonic oscillator (case LI) and to the three-dimensional Coulomb potential (case LIII) and will be discussed in the next section. We can observe the behavior of the eigenfunctions inside their respective effective potentials $V_{\rm eff}^{(I,III)}(x)$. The multiplicity of the effective potentials in the case LIII is due essentially to the dependence of $V_{\rm eff}^{(III)}(x)$ on $n$, in contrary of $V_{\rm eff}^{(I)}(x)$. The effective potential in the case LI looks like a typical {\it well} while it behaves like a {\it barrier} in the case LIII, which will bound the motion of particle. In both cases, a common behavior emerges in the sense that the depth of the effective potentials increase with increasing $l$ and the width decreases with increasing the mass-parameter $\alpha$. As a consequence of this, the eigenfunctions overlap and their associated energy-spectrum levels are equally spaced in the case LI and become closer and unequally spaced in the case LIII. \\
\indent At this stage, we start to evaluate the Wigner distribution functions for PDEM corresponding to \eref{3.2} and \eref{3.3}.
\paragraph{Case LI.}%
\noindent In order to find WDF for \eref{3.2}, we start by substituting \eref{3.1} and \eref{3.2} into \eref{1.1} and use the change of variable $\xi=\rme^{\alpha y/2}$, so
\begin{eqnarray}\label{3.4}
\fl  \mathcal W\left(\psi^{(I)}_n\big|x,p\right)&= \frac{n!}{\pi\Gamma(n+l+\frac{3}{2})}\,\mu^{2l+3}(x)&\int_0^{+\infty}\rme^{-\frac{\mu^2(x)}{2}(\xi+1/\xi)}\,\xi^{-2\rmi p/\alpha-1}\,L_n^{(l+1/2)}\left(\xi\mu^2(x)\right)\nonumber\\
                                          &&\times L_n^{(l+1/2)}\left(\xi^{-1}\mu^2(x)\right)\,\rmd\xi.
\end{eqnarray}
\indent Expanding the generalized Laguerre polynomials in their series form \cite{53}, i.e.
\[
L_n^{(a)}(x)=\sum_{k=0}^{n}(-1)^k{n+a\choose n-k}\frac{x^k}{k!},
\]
we can finally integrate \eref{3.4}, using {\bf 3.471}.9 of \cite{53}, to get WDF for the case LI which is given by
\begin{eqnarray}\label{3.5}
\fl  \mathcal W\left(\psi^{(I)}_n\big|x,p\right) &=& \frac{2\,n!}{\pi\,\Gamma\left(n+l+\frac{3}{2}\right)}\sum_{l_1=0}^n\sum_{l_2=0}^n\frac{(-1)^{l_1+l_2}}{l_1!\,l_2!}{n+l+\frac{1}{2}\choose n-l_1}{n+l+\frac{1}{2}\choose n-l_2}\nonumber \\
                                     &&\times\left(\mu^2(x)\right)^{l+l_1+l_2+3/2}\,K_{l_1-l_2-2\rmi p/\alpha}\left(\mu^2(x)\right),
\end{eqnarray}
where $K_\nu(\bullet)$ are the modified Bessel functions of the third kind and known as MacDonald function.
\paragraph{Case LIII.}%
\noindent Now we consider the eigenfunctions \eref{3.3} and following the same steps as before, one ends up by showing that WDF for the case LIII have a following expression
\begin{eqnarray}\label{3.6}
\fl  \mathcal W\left(\psi^{(III)}_n\big|x,p\right) &=& \frac{n!}{\pi\,(n+l+1)\Gamma(n+2l+2)}\sum_{l_1=0}^n\sum_{l_2=0}^n\frac{(-1)^{l_1+l_2}}{l_1!\,l_2!}{n+2l+1\choose n-l_1}{n+2l+1\choose n-l_2}\nonumber \\
                                     &&\times\left(2\,\mu(x)\right)^{2l+l_1+l_2+3}\,K_{l_1-l_2-4\rmi p/\alpha}\left(2\,\mu(x)\right).
\end{eqnarray}

\begin{figure}
 \centering
 \includegraphics[width=16cm,height=3.5cm]{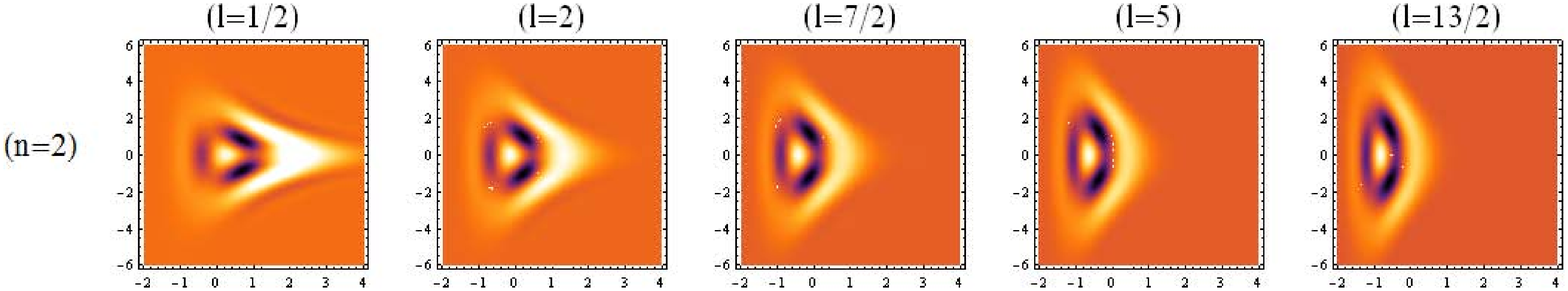}\\
 \includegraphics[width=16cm,height=3.5cm]{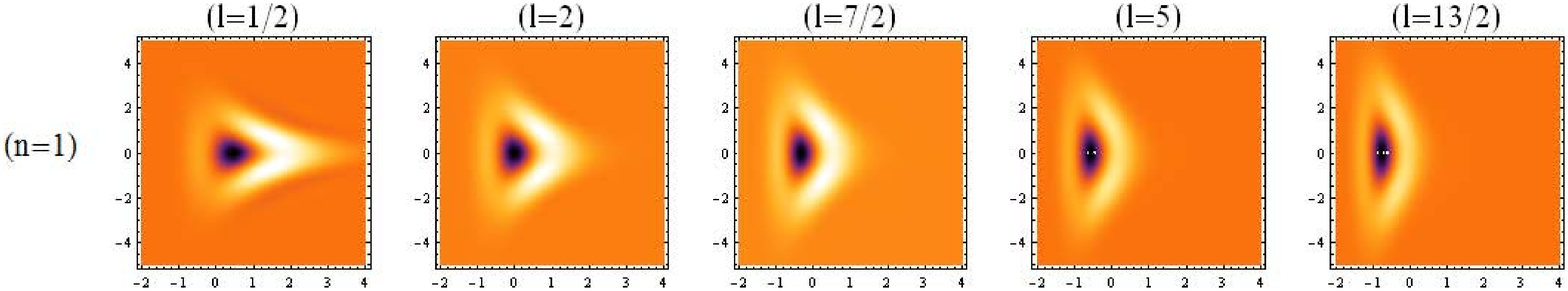}\\
 \includegraphics[width=16cm,height=3.5cm]{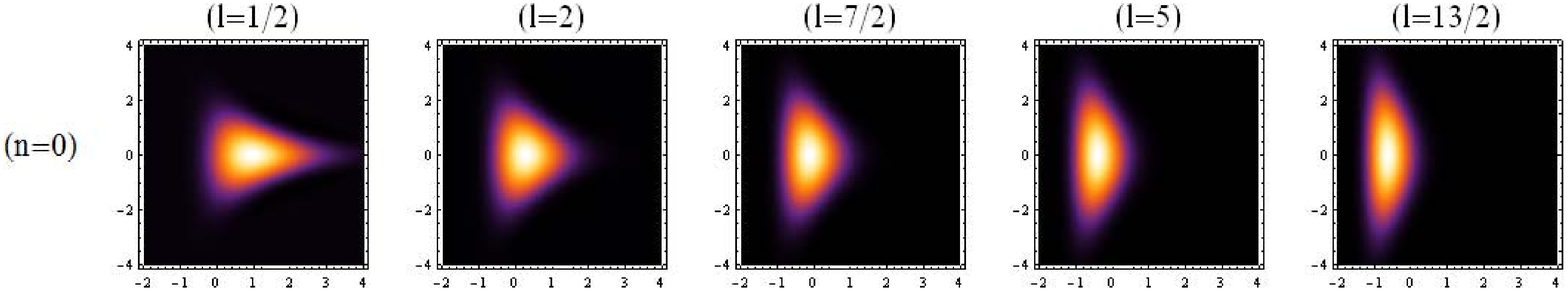}\\
 \caption{WDF of \eref{3.5} plotted for $l=\frac{1}{2},2,\frac{7}{2},5,\frac{13}{2}$, $\alpha=1$ and $n=0,1,2$. Darkness displays the minimum value while brightness designs the maximum value. The position coordinate is represented in the horizontal axis and momentum coordinate is on the vertical axis.}\label{figure2}
\end{figure}
\begin{figure}
 \centering
 \includegraphics[width=16cm,height=3.5cm]{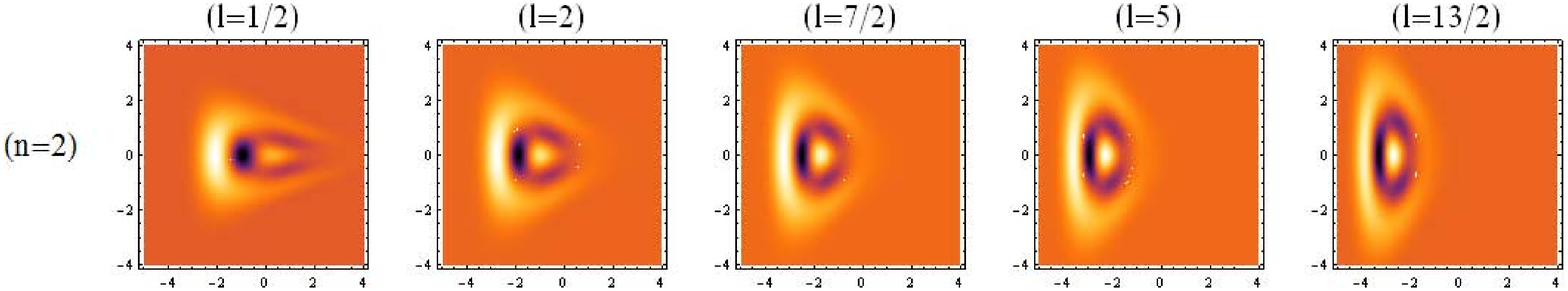}\\
 \includegraphics[width=16cm,height=3.5cm]{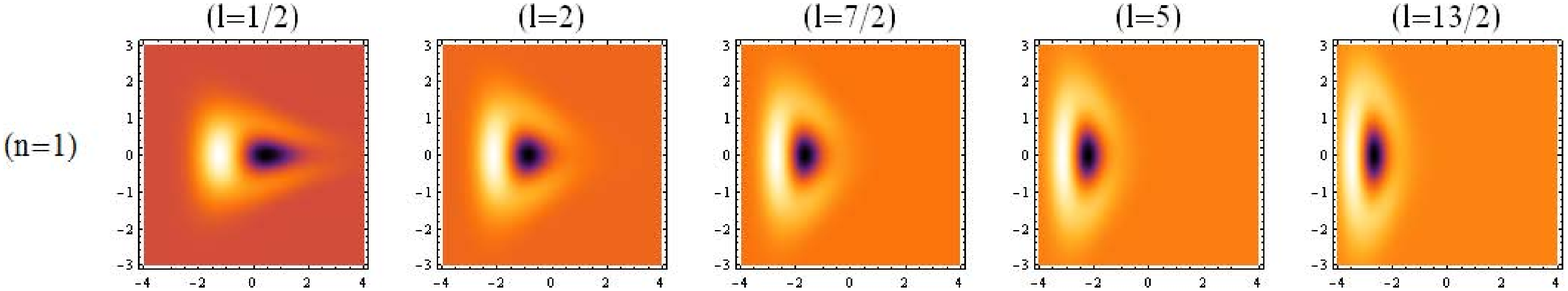}\\
 \includegraphics[width=16cm,height=3.5cm]{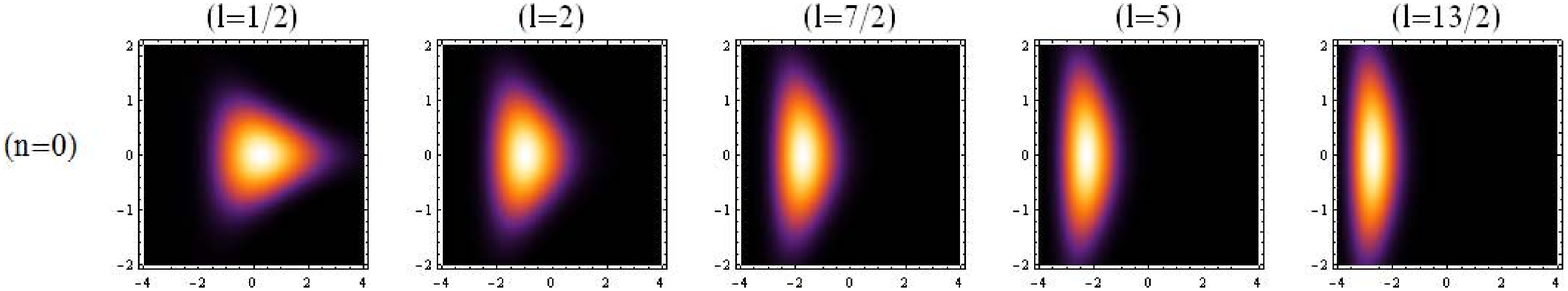}\\
 \caption{WDF of \eref{3.6} plotted for $l=\frac{1}{2},2,\frac{7}{2},5,\frac{13}{2}$, $\alpha=1$ and $n=0,1,2$. Darkness displays the minimum value while brightness designs the maximum value. The position coordinate is represented in the horizontal axis and momentum coordinate is on the vertical axis.}\label{figure3}
\end{figure}
\indent We illustrate the behavior of both distributions by plotting, in \Fref{figure2} and \Fref{figure3}, the WDF \eref{3.5} and \eref{3.6} as a function of $x$ and $p$ for $l=\frac{1}{2},2,\frac{7}{2},5,\frac{13}{2}$ and $\alpha=1$ for the ground and the two first excited states ($n=0,1,2$). One can observe that both distributions have a typical triangular shapes common to all frames. We see also that both distributions obey the inequality $|W(\psi_n(x)|x,p)|\leq\frac{1}{\pi}$. With the presence of the mass function \eref{3.1}, we can see that the dependence of the mass on the position $x$ affects WDF. Both WDF of the ground-state have, approximatively, a same deformed-Gaussian shape (as all ground states should be) with a minimum localized on $(x,p)$-plane. While for the excited states there are some regions where its values are negatives with a higher concentration. Gradually as the quantum number $l$ increases, the distributions are compressed in the $x$-direction and become more and more stretched in the direction of momentum $p$ with a slight shift to the left. However, some differences appear between our distribution shapes and those analyzed in \cite{48,49}. In our opinion this is due in two-fold: firstly, to the BenDaniel-Duke ordering prescription used by us which is different from those proposed by the authors in \cite{48,49} and secondly, to the second kind of CT applied in this paper in contrary to \cite{48,49} that have chosen to work under the first one.\\
\indent It is worth to note that in the case LI if $l$ is non negative integer and regarded as the angular momentum quantum number, then the distributions in \Fref{figure2} are associated to the three-dimensional harmonic oscillator potential. While in the case LIII, with the same features for $l$ and setting $b\equiv n+l+1=Z$ ($Z$ being the charge number), the corresponding distributions in \Fref{figure3} are those of the three-dimensional Coulomb potential.

\section{Quantum canonical and Weyl transforms for the generalized Laguerre PDEM}%

\noindent The choice of generalized coordinates in the framework of PDEM depends in general on the coordinates ($\mu$, $\pi$) adapted for better describing a problem instead of the usual coordinates ($x$, $p$) (see, for example \cite{1}). This means that in the phase space, one could find a canonical transformation from the set ($x,p$) to a new set ($\mu, \pi$) which preserves the Poisson brackets \cite{54}
\begin{eqnarray}\label{4.1}
\left\{\mu(x),\pi(x,p)\right\}_{\rm{P.B}}\equiv\frac{\partial\mu(x)}{\partial x}\frac{\partial\pi(x,p)}{\partial p}-\frac{\partial\pi(x,p)}{\partial x}\frac{\partial\mu(x)}{\partial p}=1.
\end{eqnarray}
\indent Of course, the choice of canonical transformations $(x,p)\rightarrow(\mu,\pi)$ will be indicated by the characteristics of our problem. In doing so, let us consider that $\mu$ is purely a function of the spatial coordinate, $\mu\equiv\mu(x)$, and $\pi$ is the two-variable function such that $\pi\equiv\pi(x,p)=m^\beta(x)\,p$. Then using \eref{4.1} and combining with \eref{2.13}, we get $\beta=-\frac{1}{2}$.\\
\indent Now because we are interested to evaluate a spread in position and momentum, determined by $\Delta\Theta:=\sqrt{\langle\Theta^2\rangle-\langle\Theta\rangle^2}$\,(with $\Theta=\mu,\pi$), in order to verify the Heisenberg uncertainty principle, we need to use the Weyl transformation defined in \eref{1.2} and \eref{1.3} which quantizes classical coordinates ($\mu,\pi$) to its corresponding quantum operators ($\hat\mu,\hat\pi$) (see, for example \cite{32,33,34,42}.)\\
\indent For this end, quantum canonical transformations (QCT) are regarded as a suitable transforms to find a such correspondence. Indeed QCT are defined to change the phase space variables preserving the Dirac brackets
\begin{eqnarray}\label{4.2}
  [\hat x,\hat p]_{\rm Dirac} = \rmi \equiv [\mu(\hat x),\pi(\hat x,\hat p)],
\end{eqnarray}
which are implemented by a complex function $\mathcal C(x,p)$ such that $\mu=\mathcal C\,x\,\mathcal C^{-1}$ and $\pi=\mathcal C\,p\,\mathcal C^{-1}$. For more details we refer the readers who are interested to \cite{51}. For considerations discussed above, it is convenient to use point canonical transformation implemented by the change of variable (see (56) in \cite{51})
\begin{eqnarray}
  \hat x &\rightarrow& \mu(\hat x)\quad=\mathcal P_{\mu(x)}\,\hat x\,\mathcal P_{\mu^{-1}(x)}, \label{4.3}\\
  \hat p &\rightarrow& \pi(\hat x,\hat p)=\mathcal P_{\mu(x)}\,\hat p\,\mathcal P_{\mu^{-1}(x)}\equiv\frac{1}{\left(\frac{\rmd\mu}{\rmd x}\right)}\,\hat p
                                          =\frac{1}{\sqrt{m(\hat x)}}\,\hat p. \label{4.4}
\end{eqnarray}
\indent We see that \eref{4.4} presents a convenient approach to deal with a normal ordering that has $\hat p$ to the right and it is easy to check that \eref{4.3} and \eref{4.4} satisfy the commutation relation \eref{4.2}. We are now able to calculate the Weyl transformation (WT) of $\mu^m(\hat x)$ and $\pi^m(\hat x,\hat p)$ ($m=1,2$). Note that since $\mu^m(\hat x)$ is purely a function of $\hat x$, then its WT is just the original function  with $\hat x$ replaced by $x$. However, WT of $\pi^m(\hat x,\hat p)$ will not be simply performed because they involve cross terms on $\hat x$ and $\hat p$.\\
\indent According to the definition of WT given at \eref{1.3} and after lengthly but straightforward algebra, we find after carrying out integrations that
\begin{eqnarray}
  \mathfrak W[\mu^m(\hat x)]        &=& \mu^m(x),\qquad(m=1,2),\label{4.5} \\
  \mathfrak W[\pi(\hat x,\hat p)]   &=& \frac{2}{\alpha}\,\mu^{-1}(x)\,p+\frac{\rmi}{2}\,\mu^{-1}(x),\label{4.6} \\
  \mathfrak W[\pi^2(\hat x,\hat p)] &=& \left(\frac{2}{\alpha}\,\mu^{-1}(x)\,p\right)^2+\frac{2\rmi}{\alpha}\,\mu^{-2}(x)\,p\nonumber \\
                                    &\equiv& \mathfrak{W}^2[\pi(\hat x,\hat p)]+\frac{1}{4}\,\mu^{-2}(x), \label{4.7}
\end{eqnarray}
where details and proofs of derivation of \eref{4.7} are left for the appendix A. Inserting the expressions of WDF given at \eref{3.5} and \eref{3.6} as well as \eref{4.5}-\eref{4.7} into \eref{1.2}, the analytical expressions for the expectation values $\langle\mu^m(x)\rangle$ and $\langle\pi^m(x,p)\rangle$ in the cases LI and LIII are given as follows
\paragraph{Case LI.}%
\noindent For a state represented by WDF \eref{3.5}, we get
\numparts
\begin{eqnarray}\label{4.8}
\fl  \langle\mu^m\rangle_I &=& \sum_{l_1=0}^n\sum_{l_2=0}^n\gamma^{(I)}_{n,l,l_1,l_2}\,\Gamma\left(l+l_1+l_2+\frac{m+3}{2}\right),\qquad(m=1,2)\label{4.8a}\\
\fl  \langle\pi\rangle_I   &=& \rmi\sum_{l_1=0}^n\sum_{l_2=0}^n\gamma^{(I)}_{n,l,l_1,l_2}\,\left(l_1-l_2+\frac{1}{2}\right)\,\Gamma(l+l_1+l_2+1),\label{4.8b}\\
\fl  \langle\pi^2\rangle_I &=& \sum_{l_1=0}^n\sum_{l_2=0}^n\gamma^{(I)}_{n,l,l_1,l_2}\,\left(l+\frac{1}{2}-l_1^2-l_2^2+2l_1l_2+2l_1\right)
                               \,\Gamma\left(l+l_1+l_2+\frac{1}{2}\right)\label{4.8c},
\end{eqnarray}
\endnumparts
where
\begin{eqnarray}\label{4.9}
  \gamma^{(I)}_{n,l,l_1,l_2}&=&\frac{n!}{\Gamma\left(n+l+\frac{3}{2}\right)}\,\frac{(-1)^{l_1+l_2}}{l_1!\,l_2!}\,{n+l+\frac{1}{2}\choose n-l_1}\,{n+l+\frac{1}{2}\choose n-l_2}.
\end{eqnarray}
\indent As for \eref{4.7}, the derivation of \eref{4.8a} and \eref{4.8c} are lengthly but straightforward and are kept for the appendix B.

\paragraph{Case LIII.}%
\noindent For a state represented by WDF \eref{3.6}, we have
\numparts
\begin{eqnarray}\label{4.10}
\fl  \langle\mu^m\rangle_{III} &=& \frac{1}{2^{m+1}}\sum_{l_1=0}^n\sum_{l_2=0}^n\gamma^{(III)}_{n,l,l_1,l_2}\,\Gamma(2l+l_1+l_2+m+3),\qquad(m=1,2)\label{4.10a}\\
\fl  \langle\pi\rangle_{III}   &=& \frac{\rmi}{2}\sum_{l_1=0}^n\sum_{l_2=0}^n\gamma^{(III)}_{n,l,l_1,l_2}\,(l_1-l_2+1)\,\Gamma(2l+l_1+l_2+2),\label{4.10b}\\
\fl  \langle\pi^2\rangle_{III} &=& \frac{1}{2}\sum_{l_1=0}^n\sum_{l_2=0}^n\gamma^{(III)}_{n,l,l_1,l_2}\left(2l+1-l_1^2-l_2^2-l_1+3l_2+2l_1l_2\right)
                                \Gamma(2l+l_1+l_2+1)\label{4.10c},
\end{eqnarray}
\endnumparts
with
\begin{eqnarray}\label{4.11}
\fl  \gamma^{(III)}_{n,l,l_1,l_2}&=&\frac{n!}{(n+l+1)\,\Gamma(n+2l+2)}\,\frac{(-1)^{l_1+l_2}}{l_1!\,l_2!}\,{n+2l+1\choose n-l_1}\,{n+2l+1\choose n-l_2}.
\end{eqnarray}

\begin{figure}[h]
 \centering
 \includegraphics[width=16cm,height=8cm]{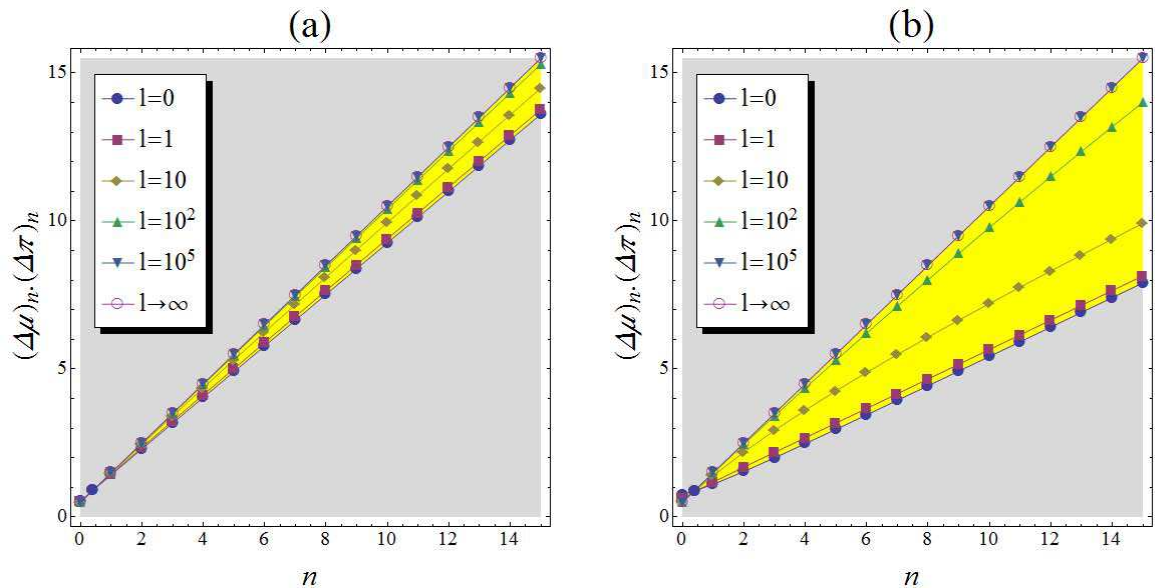}
 \caption{Uncertainty principle distributions for the angular momentum quantum number $l=0,1,10,10^2,10^5$, in comparison with the limit $l\rightarrow\infty$ given by \eref{4.12}: (a) the three-dimensional harmonic oscillator potential and (b) the three-dimensional Coulomb potential. The shaded regions correspond to the forbidden areas for uncertainty values.} \label{figure4}
\end{figure}

\indent All the expectation values deduced in (4.8) and (4.10) depend on quantum numbers $n$ and $l$. We will use them to determine the lower bound on the product of variances in the measurement of observables corresponding to canonical operators \eref{4.3} and \eref{4.4}. The exact values of the product $(\Delta\mu)_{n,l}\cdot(\Delta\pi)_{n,l}$ in both cases are computed for the nine excited states, including the ground-state, and they are reported below in \Tref{table1}.\\
\indent For the ground-state level ($n=0$), it can be straightforwardly seen that the uncertainty principle is bounded  $\frac{1}{2}<(\Delta\mu)_{0,l}\cdot(\Delta\pi)_{0,l}\lesssim\frac{3}{4}$, suggesting that $(\Delta\mu)_{0,l}\cdot(\Delta\pi)_{0,l}$ is \textit{almost} at its minimum. For instance, when $l=10^5\gg1$, one can observe in both cases that $(\Delta\mu)_{0,l}\cdot(\Delta\pi)_{0,l}$ approaches the value $\frac{1}{2}$. Therefore we expect that in the limit $l\rightarrow\infty$, a common pattern emerges as a linear-behavior
\begin{eqnarray}\label{4.12}
  (\Delta\mu)_{n,\infty}\cdot(\Delta\pi)_{n,\infty} &\equiv&\Delta_n \simeq n+\frac{1}{2},
\end{eqnarray}
applies for all levels $n=0,1,2,\cdots$.\\
\indent As an application, we illustrate in \Fref{figure4} the uncertainty principle distributions to the three-dimension harmonic oscillator potential (case LI) and to the three-dimensional Coulomb potential (case LIII), for five integer quantum numbers $l=0,1,10,10^2,10^5$, and for even and odd $n$ up to 15, in comparison with the linear distribution deduced in \eref{4.12}. We can observe that the fundamental domain for uncertainties is the area between the shaded regions. These uncertainties are confined into a narrow region (triangular yellow band) for the three-dimensional harmonic oscillator, comparing to the three-dimensional Coulomb potential. They change progressively as $l$ increases and, in the end approach \eref{4.12} when $l\rightarrow\infty$. For smaller $n$, the best $\Delta_n$-values producing the fastest convergence to \eref{4.12} are those concerned by the three-dimensional harmonic oscillator. The associated $\Delta_n$-curves are shown as solid lines in \Fref{figure4} and delimited by a forbidden regions corresponding to the larger and lower envelopes of the extremal uncertainties. Since one approaches a classical state for large quantum numbers, then we can say that the {\it quantum-classical connection} is established.

\begin{table}
\caption{\label{table1} Values of the Uncertainty principle $(\Delta\mu)_{n,l}\cdot(\Delta\pi)_{n,l}$ for different values of the angular momentum quantum number $l$ and for even and odd n up to 8. We observe that for $l=10^5$, the corresponding values approach \eref{4.12}. For an integer quantum number $l$, the case LI is associated to the three-dimensional harmonic oscillator, while the case LIII corresponds to the three-dimensional Coulomb potential.}
\footnotesize\rm
\centering
\lineup
\begin{tabular}{@{}*{11}{l}}
\br
&&\centre{9}{$n$}\\
\ns
&&\crule{9}\\
$l$ & Case & 0 & 1 & 2 & 3 & 4 & 5 & 6 & 7 & 8  \\
\mr
\tab 0              &LI & 0.546754 & 1.43040 & 2.30175 & 3.17156 & 4.04111 & 4.91069 & 5.78034 & 6.65008 & 7.51991 \\
                    &LIII& 0.749999 & 1.10397 & 1.53884 & 2.00195 & 2.47840 & 2.96214 & 3.45026 & 3.94121 & 4.43410 \\
\mr
\tab{$\frac{1}{2}$} &LI & 0.527801 & 1.44084 & 2.33128 & 3.21362 & 4.09202 & 4.96819 & 5.84297 & 6.71682 & 7.59002 \\
                    &LIII& 0.666667 & 1.14262 & 1.61521 & 2.09513 & 2.58057 & 3.06972 & 3.56137 & 4.05477 & 4.54943 \\
\mr
\tab 1              &LI & 0.519534 & 1.44954 & 2.35391 & 3.24671 & 4.13316 & 5.01570 & 5.89566 & 6.77379 & 7.65061 \\
                    &LIII& 0.625000 & 1.18014 & 1.68605 & 2.18268 & 2.67767 & 3.17281 & 3.66850 & 4.16477 & 4.66156 \\
\mr
\tab{$\frac{3}{2}$} &LI & 0.514988 & 1.45625 & 2.37137 & 3.27307 & 4.16679 & 5.05536 & 5.94036 & 6.82279 & 7.70332 \\
                    &LIII& 0.599999 & 1.21269 & 1.74938 & 2.26327 & 2.76875 & 3.27074 & 3.77116 & 4.27086 & 4.77022 \\
\mr
\tab 2              &LI & 0.512135 & 1.46147 & 2.38517 & 3.29448 & 4.19474 & 5.08891 & 5.97875 & 6.86540 & 7.74960 \\
                    &LIII& 0.583333 & 1.24019 & 1.80524 & 2.33667 & 2.85350 & 3.36317 & 3.86905 & 4.37275 & 4.87516 \\
\mr
\tab{$\frac{5}{2}$} &LI & 0.510185 & 1.46561 & 2.39632 & 3.31220 & 4.21833 & 5.11769 & 6.01210 & 6.90280 & 7.79061 \\
                    &LIII& 0.571428 & 1.26336 & 1.85433 & 2.40324 & 2.93198 & 3.45006 & 3.96204 & 4.47033 & 4.97626 \\
\mr
\tab 3              &LI & 0.508770 & 1.46897 & 2.40551 & 3.32710 & 4.23851 & 5.14264 & 6.04136 & 6.93592 & 7.82723 \\
                    &LIII& 0.562500 & 1.28300 & 1.89754 & 2.46352 & 3.00451 & 3.53151 & 4.05016 & 4.56354 & 5.07346 \\
\mr
\tab{$\frac{7}{2}$} &LI & 0.507698 & 1.47174 & 2.41321 & 3.33981 & 4.25597 & 5.16450 & 6.06724 & 6.96547 & 7.86014 \\
                    &LIII& 0.555555 & 1.29978 & 1.93572 & 2.51816 & 3.07148 & 3.60778 & 4.13354 & 4.65245 & 5.16674 \\
\mr
\tab 4              &LI & 0.506858 & 1.47406 & 2.41976 & 3.35077 & 4.27123 & 5.18380 & 6.09030 & 6.99202 & 7.88990 \\
                    &LIII& 0.550000 & 1.31424 & 1.96960 & 2.56778 & 3.13336 & 3.67916 & 4.21235 & 4.73715 & 5.25617 \\
\mr
\tab{$\frac{9}{2}$} &LI & 0.506183 & 1.47603 & 2.42539 & 3.36033 & 4.28468 & 5.20098 & 6.11100 & 7.01601 & 7.91694 \\
                    &LIII& 0.545454 & 1.32681 & 1.99982 & 2.61294 & 3.19057 & 3.74597 & 4.28682 & 4.81779 & 5.34184 \\
\mr
\tab 5              &LI & 0.505629 & 1.47773 & 2.43028 & 3.36873 & 4.29663 & 5.21638 & 6.12968 & 7.03779 & 7.94165 \\
                    &LIII& 0.541666 & 1.33783 & 2.02690 & 2.65417 & 3.24355 & 3.80853 & 4.35719 & 4.89455 & 5.42386 \\
\mr
\tab{10}             &LI & 0.502964 & 1.48698 & 2.45789 & 3.41801 & 4.36910 & 5.31253 & 6.24942 & 7.18066 & 8.10699 \\
                    &LIII& 0.522727 & 1.40126 & 2.19447 & 2.92575 & 3.61106 & 4.26162 & 4.88548 & 5.48851 & 6.07511 \\
\mr
\tab{$10^2$}        &LI & 0.500310 & 1.49846 & 2.49482 & 3.48941 & 4.48229 & 5.47348 & 6.46302 & 7.45096 & 8.43732 \\
                    &LIII& 0.502475 & 1.48781 & 2.45906 & 3.41674 & 4.36135 & 5.29337 & 6.21326 & 7.12145 & 8.01836 \\
\mr
\tab{$10^3$}        &LI & 0.500031 & 1.49984 & 2.49947 & 3.49890 & 4.49816 & 5.49723 & 6.49611 & 7.49481 & 8.49332 \\
                    &LIII& 0.500249 & 1.49875 & 2.49576 & 3.49129 & 4.48534 & 5.47791 & 6.46902 & 7.45867 & 8.44686 \\
\mr
\tab{$10^4$}        &LI & 0.500003 & 1.49998 & 2.49995 & 3.49989 & 4.49982 & 5.49972 & 6.49961 & 7.49948 & 8.49933 \\
                    &LIII& 0.500025 & 1.49988 & 2.49958 & 3.49913 & 4.49853 & 5.49778 & 6.49688 & 7.49583 & 8.49463 \\
\mr
\tab{$10^5$}        &LI & 0.5000003 & 1.49999 & 2.49999 & 3.49999 & 4.49998 & 5.49997 & 6.49996 & 7.49995 & 8.49993 \\
                    &LIII& 0.5000024 & 1.49999 & 2.49996 & 3.49991 & 4.49985 & 5.49978 & 6.49969 & 7.49958 & 8.49946 \\
\br
\end{tabular}
\end{table}

\section{Conclusion}%

\noindent In this paper, we calculated analytically and numerically the Wigner distribution functions (WDF) for the generalized Laguerre polynomials in two different cases, using an exponentially decaying mass function. Our main aim is to see how the presence of dependence of the mass function on the position-coordinate $x$ can affect WDF and preserve the Heisenberg's uncertainty principle. Using a different ordering prescription than those used in \cite{48,49}, we found that WDF have a typical triangle shapes different from those obtained by the authors. We observed that WDF are compressed in the $x$-direction and become more and more stretched in the $p$-direction as $l$ increases. We agree with \cite{48,49} by saying that this behavior indicates an apparent universality of WDF, no matter what the explicit forms of the effective-mass function $m(x)$ are. Finally we introduced an adequate quantum canonical transformation variables in order to verify the universality of the Heisenberg uncertainty principle and we used to this end the Weyl transformation to evaluate a spread in position and momentum. An interesting observation which can be made is that there is a common pattern which emerges, when $l\rightarrow\infty$, offering for the measurement of observables a lower bound, and thus preserving the Heisenberg uncertainty principle. We have found that the quantum-classical connection has been established. \\
\indent This work can be extended to consider the two other possible solutions described by eigenfunctions \eref{2.7} with different profiles for the effective mass function $m(x)$ and expressed in terms of Hermite- and Jacobi-polynomials. Quantum systems with many dimensions can also be considered since the Wigner distribution function and Weyl transformation can be generalized to D-dimensions (see, for example \cite{37}).


\appendix%

\setcounter{section}{1}

\section*{Appendix A: Derivations of \eref{4.7}, \eref{4.8a} and \eref{4.8c}}%

\noindent In this appendix, the derivation of \eref{4.7}, \eref{4.8a} and \eref{4.8c} will be provided leaving \eref{4.6} and \eref{4.8b}, which can be considered as a particular cases, to the readers. The expectation values \eref{4.10a}-\eref{4.10c} for the case LIII can be deduced in the same manner.
\paragraph{Equation \eref{4.7}.}%
To this end, let us setting $U(x)=m^{-1/2}(x)$. From the definition of the Weyl transformation \eref{1.3}, we have ($\hbar=m_0=1$)
\begin{eqnarray}\label{A.1}
\fl  \mathfrak{W}[\pi^2(\hat x,\hat p)] &=& \int_{-\infty}^{+\infty}\rme^{-\rmi py}\,\left\langle x+\frac{y}{2}
  \bigg|U(\hat x)\hat p\,U(\hat x)\hat p\bigg|x-\frac{y}{2}\right\rangle\rmd y\nonumber \\
\fl   &=& \int_{-\infty}^{+\infty}\rme^{-\rmi py}\,U^2\left(x+\frac{y}{2}\right)\left\langle x+\frac{y}{2}
  \bigg|\hat p^2\bigg|x-\frac{y}{2}\right\rangle\rmd y\nonumber \\
\fl  && -\rmi\int_{-\infty}^{+\infty}\rme^{-\rmi py}\,U\left(x+\frac{y}{2}\right)\frac{\rmd}{\rmd x}U\left(x+\frac{y}{2}\right)\left\langle x+\frac{y}{2}\bigg|\hat p\bigg|x-\frac{y}{2}\right\rangle\rmd y,
\end{eqnarray}
where we have used the commutation relation $\left[U(x),p\right]=\rmi U'(x)$. By inserting the identity operator $\hat\mathbf 1=\int|p\rangle\langle p|\,\rmd p$ on the right of $\hat p^2$ and $\hat p$, taking into account the definition $\langle x|p\rangle=(2\pi)^{-1/2}\exp(\rmi xp)$, \eref{A.1} can be expressed as
\begin{eqnarray}\label{A.2}
\fl  \mathfrak{W}[\pi^2(\hat x,\hat p)] &=& \frac{1}{2\pi}\int_{-\infty}^{+\infty}\rmd p'\int_{-\infty}^{+\infty}\rmd y\,\rme^{-\rmi py}\,U^2\left(x+\frac{y}{2}\right)p'^2\,\rme^{\rmi p'y}\nonumber \\
\fl&&-\frac{\rmi}{2\pi}\int_{-\infty}^{+\infty}\rmd p''\int_{-\infty}^{+\infty}\rmd y\,\rme^{-\rmi py}\,U\left(x+\frac{y}{2}\right)\frac{\rmd}{\rmd x}U\left(x+\frac{y}{2}\right)p''\,\rme^{\rmi p''y}.
\end{eqnarray}
\indent Next substituting $p'^2\,\rme^{\rmi p'y}$ by $-\frac{\rmd^2}{\rmd y^2}\,\rme^{\rmi p'y}$ and $p''\,\rme^{\rmi p''y}$ by $-\rmi\frac{\rmd}{\rmd y}\,\rme^{\rmi p''y}$, and the use of $\int\exp(\rmi py)\,\rmd y=2\pi\delta(p)$ lead us to carry out the $(p',p'')$-integrations, giving the derivatives of the delta function
\begin{eqnarray}\label{A.3}
\fl  \mathfrak{W}[\pi^2(\hat x,\hat p)] &=& -\frac{1}{4}\,\rme^{\alpha x}\int_{-\infty}^{+\infty}\rme^{(\alpha-2\rmi p)\varphi}\,\rmd\left(\frac{\rmd\delta(\varphi)}{\rmd\varphi}\right)-\frac{\alpha}{4}\,\rme^{\alpha x}\int_{-\infty}^{+\infty}\rme^{(\alpha-2\rmi p)\varphi}\,\rmd\delta(\varphi),
\end{eqnarray}
where we have introduced the auxiliary mass function $U(x+\varphi)=\rme^{\alpha(x+\varphi)/2}$ and used the relation $\delta(\varphi)=2\delta(2\varphi)$, with $y=2\varphi$.\\
\indent Finally integrating \eref{A.3} using
\begin{eqnarray}\label{A.4}
  \int_{-\infty}^{+\infty}\delta^{(n)}(\varphi-\varphi_0)f(\varphi)\,\rmd\varphi &=& (-1)^n\,f^{(n)}(\varphi_0),\qquad(n=0,1,2,\cdots)
\end{eqnarray}
and performing derivatives with respect to $\varphi$ at $\varphi_0=0$, we obtain the desired result \eref{4.7}. A similar treatment can be performed to deduce \eref{4.6}, but in more easier way.
\paragraph{Equation \eref{4.8a}.}%
The expectation values for the Weyl transformation of the operator $\mu^m(\hat x)$, $(m=1,2)$, can be evaluated by inserting \eref{3.5} and \eref{4.5} into \eref{1.2}
\begin{eqnarray}\label{A.5}
  \langle\mu^m\rangle &\equiv& \Sigma_{n,l}^{(I)}\int_{-\infty}^{+\infty}\rmd x\int_{-\infty}^{+\infty}\rmd p\,\mu^{m+2l+2l_1+2l_2+3}
                          K_{l_1-l_2-2\rmi p/\alpha}(\mu^2)\nonumber \\
  &\stackrel{\mbox{\footnotesize ($\ast$)}}{=}& \frac{\Sigma_{n,l}^{(I)}}{\alpha}\int_{-\infty}^{+\infty}\rmd p\int_{0}^{+\infty}
     \left(\mu^2\right)^{l+l_1+l_2+\frac{m+1}{2}}K_{l_1-l_2-2\rmi p/\alpha}(\mu^2)\,\rmd\mu^2\nonumber \\
  &\stackrel{\mbox{\footnotesize ($\ast\ast$)}}{=}& 2^{l+l_1+l_2+\frac{m-1}{2}}\,\Sigma_{n,l}^{(I)}\int_{-\infty}^{+\infty}\rmd s\,
     \Gamma\left(\varsigma_1-\rmi s\right)\Gamma\left(\varsigma_2+\rmi s\right),
\end{eqnarray}
where $s=p/\alpha$, $\varsigma_{1,2}=l/2+l_{1,2}+(m+3)/4$, and
\begin{eqnarray}\label{A.6}
 \Sigma_{n,l}^{(I)} &=& \frac{2}{\pi}\sum_{l_1=0}^n\sum_{l_2=0}^n\gamma^{(I)}_{n,l,l_1,l_2},
\end{eqnarray}
with $\gamma^{(I)}_{n,l,l_1,l_2}$ is defined in \eref{4.9}. In ($\ast$) we have used \eref{3.1} in its differential form, i.e. $\rmd x=2\rmd \mu(x)/(\alpha|\mu(x)|)$ over all configuration space and ($\ast\ast$) follows after carrying out the $\mu^2$-integration using the relation {\bf 6.561}.16 of \cite{53}. Finally, inserting \eref{A.6} into \eref{A.5} and performing the $s$-integration using the integral established in \cite{55} (see \eref{B.1} in the appendix B) which depends on the Gauss hypergeometric function reduces to a polynomials of order $c=-N,\,(N\in\mathbb N)$, we end up with
\begin{eqnarray}\label{A.7}
   \langle\mu^m\rangle &= \sum_{l_1=0}^n\sum_{l_2=0}^n\gamma^{(I)}_{n,l,l_1,l_2}\,\Gamma\left(l+l_1+l_2+\frac{m+3}{2}\right),\qquad(m=1,2)
\end{eqnarray}
which are the desired expectation values \eref{4.8a}.
\paragraph{Equation \eref{4.8c}.}%
For this equation, the expectation value $\langle\pi^2\rangle$ is given by inserting \eref{3.5} and \eref{4.7} into the definition \eref{1.2}. Following a similar treatment as before, $\langle\pi^2\rangle$ can be split into two-double integrals
\begin{eqnarray}\label{A.8}
  \langle\pi^2\rangle &\equiv& \left(\frac{2}{\alpha}\right)^3\frac{\Sigma_{n,l}^{(I)}}{2}
            \int_{-\infty}^{+\infty}p^2\,\rmd p\int_{0}^{+\infty}\left(\mu^2\right)^{l+l_1+l_2-1/2}K_{l_1-l_2-2\rmi p/\alpha}(\mu^2)\,\rmd\mu^2\nonumber \\
     &&+\frac{2\rmi}{\alpha^2}\,\Sigma_{n,l}^{(I)}
            \int_{-\infty}^{+\infty}p\,\rmd p\int_{0}^{+\infty}\left(\mu^2\right)^{l+l_1+l_2-1/2}K_{l_1-l_2-2\rmi p/\alpha}(\mu^2)\,\rmd\mu^2\nonumber \\
  &\stackrel{\mbox{\footnotesize ($\ast$)}}{=}& 2^{l+l_1+l_2+\frac{1}{2}}\,\Sigma_{n,l}^{(I)}\int_{-\infty}^{+\infty}s^2\,
     \Gamma\left(\varsigma_1-\rmi s\right)\Gamma\left(\varsigma_2+\rmi s\right)\,\rmd s\nonumber \\
     &&+2^{l+l_1+l_2-\frac{1}{2}}\,\rmi\,\Sigma_{n,l}^{(I)}\int_{-\infty}^{+\infty}s\,
     \Gamma\left(\varsigma_1-\rmi s\right)\Gamma\left(\varsigma_2+\rmi s\right)\,\rmd s,
\end{eqnarray}
where $s=p/\alpha$ and $\varsigma_{1,2}=l/2+l_{1,2}+1/4$. In ($\ast$) we have performed the $\mu^2$-integration using {\bf 6.561}.16 of \cite{53} and $\Sigma_{n,l}^{(I)}$ is given through \eref{A.6}. In order to carry out the $s$-integration in \eref{A.8}, we need once again the integral established in \cite{55}.\\
\indent So, after some straightforward mathematical manipulations \eref{A.8} becomes
\begin{eqnarray}\label{A.9}
\fl \langle\pi^2\rangle &=& \pi\,\Sigma_{n,l}^{(I)}\,\lambda_l\,\Gamma\left(\sigma_l\right)
\left[{_2}F_1\left(-1,\sigma_l;\lambda_l;\frac{1}{2}\right)-2\left(\lambda_l+1\right)\,_2F_1\left(-2,\sigma_l;\lambda_l;\frac{1}{2}\right)\right],
\end{eqnarray}
where $\lambda_l=l/2+l_2+1/4$ and $\sigma_l=l+l_1+l_2+1/2$. Now substituting \eref{A.6} into \eref{A.9} and after simplifying the hypergeometric functions, we get
\begin{eqnarray}\label{A.10}
\fl  \langle\pi^2\rangle &=& \sum_{l_1=0}^n\sum_{l_2=0}^n\gamma^{(I)}_{n,l,l_1,l_2}\,\left(l+\frac{1}{2}-l_1^2-l_2^2+2l_1l_2+2l_1\right)
                               \,\Gamma\left(l+l_1+l_2+\frac{1}{2}\right),
\end{eqnarray}
which is a desired result \eref{4.8c}. A similar treatment can be performed to deduce \eref{4.8b}.

\appendix%

\setcounter{section}{2}

\section*{Appendix B}%

\noindent In the present appendix, we present the main result established in \cite{55} by means of an integral, which may be useful for calculating the expectation values (4.8) and (4.10).\\
\indent In \cite{55} we propose a method of derivation which allow us to obtain a general integrals containing a product of two gamma functions with a monomial $x^m$, with $m\in\mathbb N$. Indeed, we show that:
\begin{eqnarray}\label{B.1}
\fl \int_{-\infty}^{+\infty}x^m\Gamma(\alpha-\rmi x)\Gamma(\beta+\rmi x)\rmd x=m!\pi\,(-\rmi)^m\,2^{1-\alpha-\beta}\,\frac{\Gamma(\alpha+\beta)}{\Gamma(\beta)} \nonumber \\
  \times\sum\frac{(-1)^M}{i_1!\,i_2!\cdots i_m!}\frac{\Gamma(\beta+M)}{(1!)^{i_1}\,(2!)^{i_2}\cdots (m!)^{i_m}}\,{_2}F_1\left(-M,\alpha+\beta;\beta;\frac{1}{2}\right),
\end{eqnarray}
where the summation is over all solutions in non negative integers of the equations:
\begin{eqnarray*}
  m=\sum_{\nu=0}^m \nu\,i_\nu,\qquad{\rm and}\qquad M=\sum_{\nu=0}^m i_\nu.
\end{eqnarray*}
\indent It is obvious that the application of \eref{B.1} is not restricted to the present paper; many others possibilities follow in physics as well as in mathematics, see \cite{55}.

\section*{References}%

\end{document}